\documentclass[twocolumn,floatfix,showpacs,nofootinbib,amsmath,amssymb,aps]{revtex4}
\usepackage{amsfonts}
\usepackage{graphicx}
\usepackage{bm}
\allowdisplaybreaks
\begin{document}

\title{Analytic structure of radiation boundary kernels 
for blackhole perturbations}
\author{Stephen R. Lau}
\email{lau@phys.utb.edu}
\altaffiliation{Present address: Department of Mathematics and Statistics, 
MSC03 2150, 1 University of New Mexico, Albuquerque, New Mexico 
87131-0001}
\affiliation{Center for Gravitational Wave Astronomy,
University of Texas at Brownsville, Brownsville, Texas 78520}
\begin{abstract}
Exact outer boundary conditions for gravitational perturbations of the 
Schwarzschild metric feature integral convolution between a time--domain 
boundary kernel and each radiative mode of the perturbation. For 
both axial (Regge--Wheeler) and polar (Zerilli) perturbations, we study 
the Laplace transform of such kernels as an analytic function of 
(dimensionless) Laplace frequency. We present numerical evidence 
indicating that each such frequency--domain boundary kernel admits a 
``sum--of--poles'' representation. Our work has been inspired by Alpert, 
Greengard, and Hagstrom's analysis of nonreflecting boundary conditions 
for the ordinary scalar wave equation.
\end{abstract}
\maketitle

\section{Introduction}
Now 50 years old, the perturbation theory of Schwarz\-schild blackholes 
remains a timely subject 
with fundamental applications. The covariant d'Alembertian (or wave 
equation) associated with the Schwarzschild line--element describes 
scalar ``perturbations.'' Classical Schwarzschild blackholes are spherically 
symmetric and static in time, and these symmetries allow for combined 
multipole and Fourier (or Laplace) decompositions. As a result, 
perturbations may be described via a denumerable collection of ODE 
rather than the d'Alembertian PDE. Similar ODE describe 
electromagnetic ``perturbations'' \cite{Wheeler} and gravitational 
perturbations (small genuine fluctuations in the background geometry) 
\cite{ReggeWheeler,Zerilli}, although their derivation is more 
complicated since the multipole decomposition involves either 
vector or tensor spherical harmonics. Remarkably, via the technique 
of ``despinning'' based on the $\eth$ operator \cite{Price72b}, 
such ODE can be related to scalar wave equations.

Wheeler considered the case of electromagnetic perturbations in 1955 
\cite{Wheeler}, showing for a given multipole that each of the two 
electromagnetic polarization states are described by a copy of a 
single ODE. Regge and Wheeler then derived a similar ODE 
describing odd--parity (or axial) gravitational perturbations in 
1957 \cite{ReggeWheeler}, and Zerilli introduced an ODE 
describing even--parity (or polar) gravitational perturbations in 
1970 \cite{Zerilli}. In the 1970s Chandrasekhar and Detweiler 
demonstrated that the Zerilli equation can be derived from the 
Regge--Wheeler equation, although the derivation involves differential 
operations (see \cite{ChandraDet} and references 
therein). In their treatment \cite{AndPrice} of {\em intertwining 
operators}, Anderson and Price clarified the relationship 
between solutions to the Regge--Wheeler and Zerilli ODE. 
Application of a first--order differential operator transforms 
smooth solutions of one equation into solutions of the other.

Schwarzschild perturbation theory has played a central role in several 
modern areas of classical and quantum gravity. Although the following 
is by no means an exhaustive list, we mention four salient 
applications: the ``close--limit'' approximation for blackhole 
collisions, a time--domain approach to the radiation reaction
problem, the asymptotic form of high frequency quasinormal modes,
and quantum uncertainty in blackhole horizons. Price and Pullin,
assuming that the colliding blackholes are initially cloaked in
a common horizon, have used first--order perturbation theory and
the Zerilli equation to compute the energy radiated away by
gravitational waves \cite{PricePullin}. They initially applied
their close--limit approximation to time--symmetric initial data. 
More general data was subsequently considered \cite{NBPP}, and the
accuracy of the approximation studied via second--order
perturbation theory \cite{GNPP1,GNPP2}. Lousto has studied the
binary radiation reaction problem in the extreme mass ratio
limit via time--domain calculations \cite{Lousto1}. His approach
relies on numerical simulation of Schwarzschild perturbations,
and he has developed a fourth--order convergent algorithm for
such simulations \cite{Lousto2}. Interest in the quasinormal mode
spectrum of classical Schwarzschild blackholes has been renewed
by possible connections with the Barbero--Immirzi parameter in loop 
quantum gravity \cite{Olaf1,Baez,DomLew}. While these issues
are beyond us, we note that they have focused attention on the
asymptotic form of high frequency quasinormal modes
\cite{MotlandNeitzke,MusiriSiopsis} (a purely classical issue).
Most recently, York and Schmekel have considered a truncated 
superspace of blackhole fluctuations, using path integral 
quantization to derive a 
Schr\"{o}dinger equation and estimate the quantum uncertainty in 
the horizon \cite{YorkSchmekel}. York and Schmekel's analysis
constitutes and improved derivation of results already found
by York via semiclassical arguments.

This paper addresses the mathematical issue of {\em exact} radiation 
outer boundary conditions ROBC for Schwarz\-schild blackhole 
perturbations. Our work is based on an approach \cite{AGH}
developed by Alpert, Greengard, and Hagstrom (AGH) for nonreflecting 
boundary conditions and time--domain wave propagation on flat spacetime. 
Beyond being of theoretical interest, such boundary conditions are 
relevant for numerical simulations. Indeed, long--time simulations 
require some form of domain reduction, that is specification of 
appropriate boundary conditions at the ``edge'' of the computational 
domain. Exact outer boundary conditions for gravitational perturbations 
of Schwarzschild blackholes feature integral convolution between a 
time--domain boundary kernel and each angular mode of the 
perturbation. For both axial (Regge--Wheeler) and polar (Zerilli) 
perturbations, we study the Laplace transform of the such kernels 
as an analytic function of (dimensionless) Laplace frequency 
$\sigma$. We present numerical evidence indicating that each such 
gravitational boundary kernel admits a ``sum--of--poles'' 
representation. These representations are similar to those considered 
by AGH for wave propagation on flat 3+1 and 2+1 spacetimes. We hope 
to interest analysts in our conjectured sum--of--poles representation
and the theorems we believe are lurking behind it.

We have considered such boundary conditions in detail before 
\cite{LauROBC1,LauROBC2} (hereafter referred to as Papers I and 
II). However, this paper goes beyond these references in the 
following ways. First, we consider the Zerilli equation, not 
considered at all in Papers I and II. Second, we focus here on 
the analytic structure of gravitational kernels (both 
Regge--Wheeler and Zerilli cases), whereas Paper I almost exclusively 
considered kernels for scalar wave propagation. Although the theory 
and methods in I and II were also spelled out for the 
gravitational (spin 2) Regge--Wheeler equation, we consider 
these results in more detail here. We also discuss the asymptotic
agreement between Schwarzschild ROBC and flatspace nonreflecting
boundary conditions, remarking on some fine points unmentioned in
our earlier work. Part of this paper roughly 
parallels Section 3 from Paper I, which documented several numerical 
tests of the sum--of--poles representation for scalar (spin 0) kernels. 
Here we run through those same tests for the gravitational kernels, 
and also carry out another test based on the Argument Principle. 
A longer paper on numerical implementation of these ideas, 
complete with extensive numerical tables, is forthcoming 
\cite{LauEvans}, and in part this paper is meant to lay more 
groundwork for that longer work.

\section{Preliminaries}
\subsection{Wave equations for gravitational perturbations}
In terms of dimensionless time $\tau$ and radius $\rho > 1$, the
standard Schwarzschild line--element reads\footnote{With $M$ the 
mass parameter of the solution, standard physical coordinates are 
then $t = 2M\tau$ and $r = 2M\rho$. We also consider a dimensionless 
Laplace frequency $\sigma$ related to the physical frequency $s$ 
via $\sigma = 2M s$.}
\begin{equation}
\mathrm{d}s^2 = -F\mathrm{d}\tau^2 + F^{-1}\mathrm{d}\rho^2
+ \rho^2\big(\mathrm{d}\theta^2 +\sin^2\theta\mathrm{d}\phi^2\big),
\end{equation}
where $F(\rho) = 1-\rho^{-1}$. 
As is well known (see, for example, the review
\cite{KokSchmidt}), radiative perturbations of the Schwarzschild 
metric are ultimately described by wave equations for 
angular modes $\Psi_{\ell m}(\tau,\rho)$. From now on, we drop the 
``azimuthal'' index $m$ and write $\Psi_{\ell}(\tau,\rho)$, since 
all equations depend only on the ``orbital'' index $\ell$.
In terms of the Regge--Wheeler tortoise coordinate 
$\rho_* = \rho + \log(\rho-1)$, the perturbation $\Psi_\ell$ 
obeys the wave equation
\begin{equation}
\frac{\partial^2\Psi_\ell}{\partial\tau^2}
- \frac{\partial^2\Psi_\ell}{\partial\rho_*^2}
+ V(\rho)\Psi_\ell = 0.
\label{onedeewaveeq}
\end{equation}
For axial perturbations $V(\rho)$ is the 
Regge--Wheeler potential
\begin{equation}
V^{RW}(\rho) 
= \left(1-\frac{1}{\rho}\right)\left[\frac{\ell(\ell+1)}{\rho^2}
+ \frac{1-\jmath^2}{\rho^3}\right],
\end{equation}
where $\jmath = 2$. 
Scalar and electromagnetic perturbations correspond to $\jmath = 0$ 
and $1$, respectively. 

Eq.~(\ref{onedeewaveeq}) also describes polar 
perturbations, although in this case $V(\rho)$ is the 
Zerilli potential
\begin{eqnarray}
\lefteqn{V^{Z}(\rho)  = }
& &
\label{Zerpotential} \\
& &  \left(1-\frac{1}{\rho}\right)\left[
\frac{8 n^2(n+1)\rho^3 + 12 n^2 \rho^2 
+ 18 n\rho + 9}{\rho^3(2 n \rho + 3)^2}
\right],
\nonumber
\end{eqnarray}
where $n = \frac{1}{2}(\ell-1)(\ell+2)$. Throughout this paper 
we consider the same objects for the Regge--Wheeler and Zerilli 
cases, and we shall make use of the following convention. An 
object without a superscript letter, say $V(\rho)$, will refer 
to the generic object, and could correspond to either of the 
two cases (actually four cases, since the Regge--Wheeler 
scenario is three cases in itself). Superscript letters will 
denote the specific cases. For example, we have $V^{RW}(\rho)$ 
and $V^{Z}(\rho)$ as above. We will also sometime use a 
superscript $F$ to denote corresponding flatspace objects,
for example
\begin{equation}
V^{F}(\rho) = 
\frac{\ell (\ell+1)}{\rho^2}
\end{equation}
as the flatspace potential.

Formal Laplace transformation of the Regge--Wheeler 
wave equation (\ref{onedeewaveeq}) yields a second--order 
ODE
\begin{equation}
L^{RW} \hat{\Psi}_\ell = 0,\quad L^{RW} = 
\frac{\mathrm{d^2}}{\mathrm{d}\rho^2_*} - V^{RW}(\rho) - 
\sigma^2,
\label{RWLaplace}
\end{equation}
where $\sigma$ is Laplace frequency.
This equation is ---apart from a transformation on 
the dependent variable--- a special case of the 
{\em confluent Heun equation} \cite{CHE1,CHE2}. 
Therefore, implementation of ROBC for the Regge--Wheeler equation 
involves confluent Heun functions, whereas the flatspace 
implementation of AGH \cite{AGH} involves Bessel 
functions (closely related to {\em confluent} hypergeometric 
functions). 

Likewise, we may
consider formal Laplace transformation of the
of the Zerilli wave equation (\ref{onedeewaveeq}),
\begin{equation}
L^{Z} \hat{\Psi}_\ell = 0,\quad L^{Z} =
\frac{\mathrm{d}^2}{\mathrm{d}\rho^2_*} - V^{Z}(\rho) -
\sigma^2.
\label{ZLaplace}
\end{equation}
Solutions of (\ref{RWLaplace}) with $\jmath = 2$ are 
related to solutions of (\ref{ZLaplace}) and {\em vice versa}
by the intertwining relations 
\cite{AndPrice}
\begin{equation}
D_{+} L^{RW} = L^{Z} D_{+},\quad D_{-} L^{Z} = L^{RW}D_{-},
\end{equation}
with
\begin{equation}
D_{\pm} = \frac{\mathrm{d}}{\mathrm{d}\rho_*} \pm
\left[{\textstyle\frac{2}{3}} n(n+1) +
\frac{3(\rho-1)}{\rho^2(3 + 2 n \rho)}\right].
\end{equation}
We have mentioned that Papers I and II examined the exact 
ROBC for the Regge--Wheeler equation, although mostly
focusing on the $\jmath = 0$ case. A natural question then
is whether or not these boundary conditions can be easily
carried over to the Zerilli case via use of intertwining 
relations. It would seem the answer is ``no,'' and we have 
been unable to make direct use of our previous 
work on ROBC via the intertwining relations. Although we 
believe the issue may merit further study, in
this paper we develop and describe Zerilli ROBC from
scratch.

\subsection{Radiation outer boundary conditions}

Let us assume initial data of compact support, and that the
radial location $\rho_B$ of the outer boundary $B$ lies beyond
the support of the data. Then the exact nonlocal ROBC 
is the following differential--integral identity:\cite{LauROBC1}
\begin{eqnarray}
\lefteqn{\left.\left(\frac{\partial\Psi_\ell}{\partial\tau} + 
\frac{\partial\Psi_\ell}{\partial\rho_*}\right)\right|_{\rho = \rho_B} =}
& & \nonumber  \\
& & \frac{F(\rho_B)}{\rho_B}
\int^\tau_0 \omega_\ell(\tau-\tau';\rho_B)\Psi_\ell(\tau',\rho_B)\mathrm{d}\tau'.
\label{therobc}
\end{eqnarray}
The ROBC equates an outgoing characteristic 
derivative of the field with an integral 
convolution of the field history. Although the conditions (\ref{therobc})
are equally valid for the Regge--Wheeler and Zerilli cases, we must
consider two separate kernels: $\omega^{RW}_\ell(\tau;\rho_B)$ and
$\omega^{Z}_\ell(\tau;\rho_B)$. As a function of complex Laplace frequency 
$\sigma$, Paper I has considered the Laplace transform 
$\hat{\omega}^{RW}_\ell(\sigma;\rho_B)$ of the integral kernel 
$\omega^{RW}_\ell(\tau;\rho_B)$, in particular arguing that it 
admits a ``sum--of--poles'' representation quite similar to the 
representation of frequency--domain kernels for flatspace wave propagation 
in 2+1 dimensions. Such representations involve both a finite pole 
sum as well as a continuous sector \cite{AGH}. The analysis in Paper I
mostly concentrated on the $\hat{\omega}^{RW}_\ell(\sigma;\rho_B)$ 
relevant for scalar wave propagation (that is for $\jmath = 0$), 
and did not consider the Zerilli equation at all. Below we consider 
the analytic structure 
of both $\hat{\omega}^{RW}_\ell(\sigma;\rho_B)$ for $\jmath = 2$ and 
$\hat{\omega}^Z_\ell(\sigma;\rho_B)$ in some detail.

\begin{figure}
\includegraphics[width=0.48\textwidth]{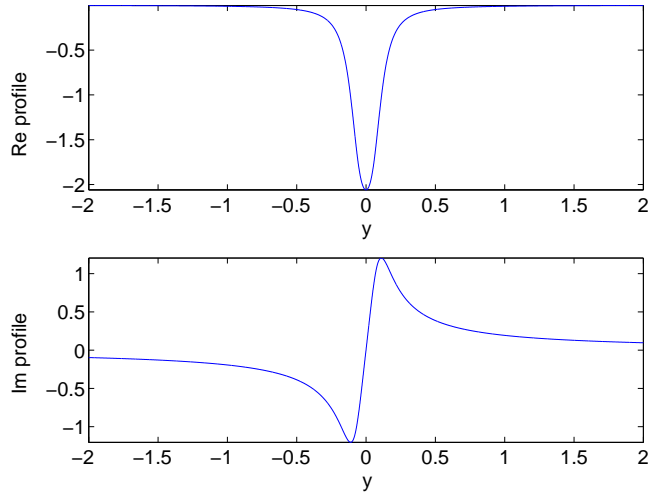}
\caption{Re$\hat{\omega}^{RW}_2(\mathrm{i}y;15)$ and 
Im$\hat{\omega}^{RW}_2(\mathrm{i}y;15)$ profiles for the 
$\jmath = 2$ 
frequency domain kernel $\hat{\omega}^{RW}_2(\mathrm{i}y;15)$. 
As indicated, $\ell = 2$ and $\rho_B = 15$.}
\label{twofigs}
\end{figure}

Paper I has developed numerical methods for evaluating 
$\hat{\omega}_\ell(\sigma;\rho_B)$ along the axis of imaginary 
Laplace frequency, precisely the contour over which the inverse 
Laplace transform is taken to obtain the time--domain kernel. We
further touch upon these methods below, but mention here that
they rely on stable numerical integration over various paths in 
both the complex $\rho$ and complex $z = \sigma\rho$ planes. 
All of our work in this paper is based upon these methods, and
we have used them to plot in {\sc FIG.}~\ref{twofigs} the 
$\jmath = 2$ Regge--Wheeler profiles 
Re$\hat{\omega}^{RW}_2(\mathrm{i}y;15)$ and 
Im$\hat{\omega}^{RW}_2(\mathrm{i}y;15)$ 
for real $y$. Although different, the corresponding Zerilli
profiles would be indistinguishable to the eye, were they also
plotted in {\sc FIG.}~\ref{twofigs}.
\vspace{0.4cm}
\begin{table}
\small
\begin{tabular}{lcc}
\hline
\\
$k$ & Re$\beta^{RW}_{2,k}(15)$ & Im$\beta^{RW}_{2,k}(15)$
 \\
\hline
 1 & $-$3.75616176922E$-$01  & 0
 \\
 2 & $-$2.52285897920E$-$01  & 0
 \\
 3 & $-$1.71458781191E$-$01  & 0
 \\
 4 & $-$1.16562490243E$-$01  & 0
 \\
 5 & $-$7.64331990276E$-$02  & 0
 \\
 6 & $-$4.68091057981E$-$02  & 0
 \\
 7 & $-$2.63730137927E$-$02  & 0
 \\
 8 & $-$1.25652994567E$-$02  & 0
 \\
 9 & $-$9.47795178947E$-$02  & 5.99312024947E$-$02
 \\
 & &
\\
\hline
\\
 $k$ & Re$\gamma^{RW}_{2,k}(15)$ & Im$\gamma^{RW}_{2,k}(15)$
 \\ \hline
 1 & $-$9.42815440764E$-$06 & 0
 \\
 2 & $-$3.66046310052E$-$04 & 0
 \\
 3 & $-$3.74027383589E$-$03 & 0
 \\
 4 & $-$8.72734265927E$-$03 & 0
 \\
 5 & $-$1.47189136342E$-$03 & 0
 \\
 6 & $-$5.01356988668E$-$05 & 0
 \\
 7 & $-$9.73423621068E$-$07 & 0
\\
 8 & $-$7.28807025058E$-$09 & 0
 \\
 9 & $-$8.94836172991E$-$02 & 6.20643548937E$-$02
\\
& &
\end{tabular}
\caption{\label{tab:tableRW}
Compressed $\jmath = 2$ Regge--Wheeler kernel for 
$\ell =  2$, $\rho_B =  15$, $\varepsilon = 10^{-10}$. There are $d = 10$
poles and strengths, and complex conjugation of the ninth
entries gives the tenth entries. Zeros correspond to outputs
from the compression algorithm which are less than $10^{-30}$
in absolute value.}
\end{table}
\begin{table}
\small
\begin{tabular}{lcc}
\hline
\\
$k$ & Re$\beta^Z_{2,k}(15)$ & Im$\beta^Z_{2,k}(15)$
 \\
\hline
 1 & $-$3.70827103177E$-$01 & 0
 \\
 2 & $-$2.48916278532E$-$01 & 0
 \\
 3 & $-$1.69097983283E$-$01 & 0
 \\
 4 & $-$1.14894340803E$-$01 & 0
 \\
 5 & $-$7.53169595130E$-$02 & 0
 \\
 6 & $-$4.61252633339E$-$02 & 0
 \\
 7 & $-$2.59880681276E$-$02 & 0
 \\
 8 & $-$1.23819599759E$-$02 & 0
 \\
 9 & $-$9.34065839850E$-$02 & 5.89802789971E$-$02 
 \\
 & &
\\
\hline
\\
 $k$ & Re$\gamma^Z_{2,k}(15)$ & Im$\gamma^Z_{2,k}(15)$
 \\ \hline
 1 & $-$8.94981130535E$-$06 & 0
 \\
 2 & $-$3.47903587670E$-$04 & 0
 \\
 3 & $-$3.56001528614E$-$03 & 0
 \\
 4 & $-$8.35011768248E$-$03 & 0
 \\
 5 & $-$1.41306545024E$-$03 & 0
 \\
 6 & $-$4.81777344134E$-$05 & 0
 \\
 7 & $-$9.35693850095E$-$07 & 0
\\
 8 & $-$7.00652022551E$-$09 & 0
 \\
 9 & $-$8.70154844197E$-$02 & 6.01803999946E$-$02
\\
& &
\end{tabular}
\caption{\label{tab:tableZ}
Compressed Zerilli kernel for $\ell =  2$, $\rho_B =  15$,
$\varepsilon = 10^{-10}$. There are $d = 10$
poles and strengths, and complex conjugation of the ninth
entries gives the tenth entries. Zeros correspond to outputs
from the compression algorithm which are less than $10^{-30}$
in absolute value.}
\end{table}
\subsection{Kernel compression}
With the ability to generate such numerical profiles for exact 
frequency--domain kernels, we have employed the technique of 
kernel compression in order to construct highly accurate numerical 
kernels which allow for efficient evaluation of the convolution 
appearing in (\ref{therobc}). Introduced by AGH \cite{AGH} and 
described further in both \cite{Jiang} and Paper II, 
compression is vital both for high--$\ell$ kernels as well as 
low--$\ell$ kernels which are dominated by costly continuous 
sectors (such sectors are further described below). The 
technique produces a rational function,
\begin{equation}
\hat{\xi}_\ell(\sigma;\rho_B) = \sum_{k=1}^d 
\frac{\gamma_{\ell,k}(\rho_B)}{\sigma - \beta_{\ell,k}(\rho_B)},
\end{equation}
which approximates $\hat{\omega}_\ell(\sigma;\rho_B)$ and 
is in fact a sum of $d$ simple poles.
The pole locations $\beta_{\ell,k}(\rho_B)$ and strengths 
$\gamma_{\ell,k}(\rho_B)$ ---output from the compression 
algorithm--- lie in the lefthalf plane. The approximation
is rigged to satisfy
\begin{equation}
\mathrm{sup}_{y\in R}\frac{\left|
\hat{\xi}_\ell(\mathrm{i}y;\rho_B)
-\hat{\omega}_\ell(\mathrm{i}y;\rho_B)\right|}{
\left|\hat{\omega}_\ell(\mathrm{i}y;\rho_B)\right|} < \varepsilon,
\end{equation}
where $\varepsilon$ is a chosen numerical tolerance. 
Theoretically, this bound on the relative supremum error in
the frequency domain ensures a long--time bound on the relative 
convolution error associated with (\ref{therobc}), 
as discussed in \cite{AGH} and Paper II. This convolution error 
arises when using the approximate time--domain kernel
\begin{equation}
\xi_\ell(\tau;\rho_B) = \sum_{k = 1}^d
\gamma_{\ell,k}(\rho_B) \exp\left[
\beta_{\ell,k}(\rho_B) \tau\right]
\end{equation}
in place of the true kernel $\omega_\ell(\tau;\rho_B)$. 
Since $\xi_\ell(\tau;\rho_B)$ is a sum of exponentials, due to 
recursive identities an approximation of the convolution 
(\ref{therobc}) based on $\xi_\ell(\tau;\rho_B)$ is not memory 
intensive. We list representative compressed kernels for
Regge--Wheeler and Zerilli kernels in Tables \ref{tab:tableRW} 
and \ref{tab:tableZ}. The two kernels are strikingly similar,
although they differ by a relative supremum error of
about $3.03\times 10^{-2}$, well below their stated 
$\varepsilon = 10^{-10}$ tolerances. Extensive numerical tables 
of compressed kernels are being prepared and will appear in  
\cite{LauEvans}.

\subsection{Quasinormal ringing and decay tail}

Providing little detail, we now carry out a simple experiment 
meant only to indicate that the described ROBC work well for
long--time simulations. More careful experiments were considered
in Paper II (for Regge--Wheeler cases only) and will be 
considered in \cite{LauEvans}.
In terms of retarded time $\mu = \tau -\rho_*$ and the pulse
function
\begin{equation}
g(\mu) = \left\{
\begin{array}{rl}
[\mu (\mu+4)]^4\big/256, & 
\quad \mathrm{for} -4\leq\mu\leq 0,\\
0, & \quad \mathrm{otherwise},
\end{array}
\right.
\end{equation}
we construct the $\ell = 2$ wave packet
\begin{equation}
\Psi_2(0,\rho) = g(-\rho_*)
\end{equation}
as initial data for the Zerilli equation. The initial packet is
then of unit height and compactly supported on $0\leq \rho_*\leq 4$.
To complete the data, we assume that
\begin{equation}
\left.\left[\frac{\partial\Psi_2}{\partial\tau}
+ \frac{\partial\Psi_2}{\partial\rho_*}
\right]\right|_{\tau = 0} = 0,
\end{equation}
so that the pulse starts as essentially outgoing. 
We place the inner boundary at
$\rho_* = -175$, and adopt (\ref{therobc}) as the boundary 
condition at $\rho = \rho_B = 15$, with an approximation of 
the exact ROBC based on the compressed kernel listed in the 
Table \ref{tab:tableZ}.

\begin{figure}
\includegraphics[width=0.47\textwidth]{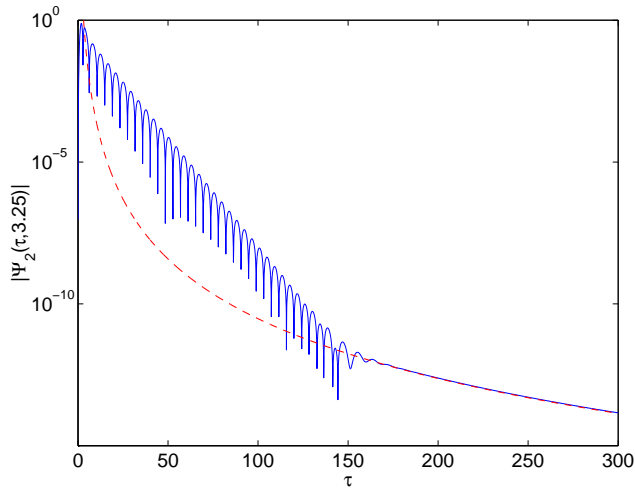}
\caption{Quasinormal ringing and decay tail.}
\label{ringtail}
\end{figure}

We evolve the data until $\tau = 300$, using the MacCormack 
predictor--corrector algorithm.\footnote{A scheme in consistent 
conservation form \cite{LeVeque} when applied to a hyperbolic 
conservation law.} Throughout the evolution,
we record the value $\Psi_2(\tau,3.25)$ of the field, that is to 
say, we record the history of the field in time at the fixed 
location $\rho = 3.25$ (actually at the grid point nearest this
location). Notice that the total run time is not long enough
for the history $\Psi_2(\tau,3.25)$ to be influenced by reflection
off of the inner boundary. The absolute value of the 
history is depicted as a 
linear--log plot in {\sc FIG.}~\ref{ringtail}. It exhibits
{\em quasinormal ringing} \cite{KokSchmidt} until about $\tau = 150$, 
and afterward a {\em Price tail} with the field decaying as 
$t^{-(2\ell+3)} = t^{-7}$ \cite{KokSchmidt}. This late--time
behavior stems from backscatter of the outgoing packet off 
of the long--range potential $V^{Z}(\rho)$. The dashed
curve $3000 t^{-7}$ has been eyeballed to fit
the late--time decay tail.

\section{Sum--of--poles representation}

\subsection{Flatspace FDNK and outgoing solution}

\begin{figure}
\scalebox{0.70}{\includegraphics{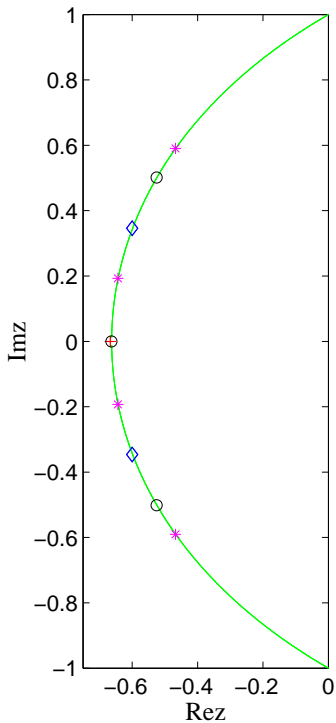}}
\caption{{\sc Scaled zeros of MacDonald functions.}
Here we plot scaled zeros $(\ell+1/2)^{-1}b_{\ell,k}$ for
$\ell = 1,2,3,4$. The cross is the scaled zero of
$K_{1/2}(z)$, the diamonds are the scaled zeros of
$K_{3/2}(z)$, the circles are the scaled zeros of
$K_{5/2}(z)$, and the stars are the scaled zeros
of $K_{7/2}(z)$.
\label{fixedcurve}}
\end{figure}

AGH have considered nonreflecting boundary conditions (NRBC) 
for both 3+1 and 2+1 flatspace wave propagation, thoroughly 
treating both theoretical description and numerical approximation 
of NRBC for both scenarios \cite{AGH}. For later comparison with the 
two gravitational scenarios, let us briefly recall the principal 
theoretical aspects of their work for the 3+1 scenario. To 
facilitate the comparison, we will use the same letters $\rho$, 
$\tau$, and $\sigma$ used for our dimensionless Schwarzschild 
coordinates, although for the flatspace case $r$, $t$, and $s$ would 
be more standard notations. Their boundary condition for a flatspace 
order--$\ell$ multipole $\Psi_\ell$ is
\begin{eqnarray}
\lefteqn{\left.\left(
\frac{\partial\Psi_\ell}{\partial\tau} + 
\frac{\partial\Psi_\ell}{\partial\rho}\right)
\right|_{\rho = \rho_B} =}
& & \nonumber  \\
& & \frac{1}{\rho_B}
\int^\tau_0 \omega^F_\ell(\tau-\tau';\rho_B)
\Psi_\ell(\tau',\rho_B)\mathrm{d}\tau',
\label{flatspacerobc}
\end{eqnarray}
comparable with (\ref{therobc}).
The exact TDRK $\omega^F_\ell(\tau;\rho_B)$ appearing in 
(\ref{flatspacerobc}) is now also a time--domain nonreflecting 
kernel (TDNK), and it is the inverse Laplace transform of an exact 
frequency--domain nonreflecting kernel (FDNK) admitting the following 
representation:\cite{AGH}
\begin{equation}
\hat{\omega}^{F}_\ell(\sigma;\rho_B) = \sum_{k=1}^\ell 
\frac{b_{\ell,k}/\rho_B}{\sigma-b_{\ell,k}/\rho_B},
\label{fdnrk}
\end{equation}
where the $b_{\ell,k}$ are the zeros of the classical MacDonald function 
$K_{\ell+1/2}(z)$, a modified cylindrical Bessel function. Here the
Bessel order is a half--integer, since we are considering the radial wave 
equation stemming from ordinary wave propagation on 3+1 flat spacetime,
and these functions have the form \cite{Watson}
\begin{eqnarray}
K_{1/2}(z) & = & \sqrt{\frac{\pi}{2z}}e^{-z},
\nonumber \\
K_{3/2}(z) & = & \sqrt{\frac{\pi}{2z}}e^{-z}
\left(1+\frac{1}{z}\right),
\nonumber \\
K_{5/2}(z) & = & \sqrt{\frac{\pi}{2z}}e^{-z}
\left(1+\frac{3}{z}+\frac{3}{z^2}\right),
\nonumber \\
K_{7/2}(z) & = & \sqrt{\frac{\pi}{2z}}e^{-z}
\left(1+\frac{6}{z}+\frac{15}{z^2}+\frac{15}{z^3}\right),
\nonumber \\
& \vdots & \\
K_{\ell+1/2}(z) & = &  \sqrt{\frac{\pi z}{2}} (-1)^\ell z^\ell
\left(\frac{1}{z}\frac{\mathrm{d}}{\mathrm{d}z}\right)^\ell
\frac{e^{-z}}{z}.
\nonumber
\end{eqnarray}
The function $K_{\ell+1/2}(z)$ has $\ell$ simple zeros 
$\{b_{\ell,k}:k=1,\cdots,\ell\}$. When scaled by order,
these zeros $(\ell+1/2)^{-1}b_{\ell,k}$ are known to 
accumulate on a fixed transcendental curve $\mathcal{C}$ in 
the lefthalf plane \cite{AGH,Olver} (the numerical methods 
developed in Paper I take advantage of this fact). Although 
this accumulation is asymptotic with large order $\ell+1/2
\rightarrow\infty$, FIG.~\ref{fixedcurve} shows that the 
agreement holds even for the lowest $\ell$, at least to the
eye. The curve $\mathcal{C}$ shown in FIG.~\ref{fixedcurve} 
has parametric form \cite{Olver,Jiang}
\begin{equation}
z(\lambda) = -\sqrt{\lambda^2-\lambda\tanh\lambda}\pm {\rm i}
\sqrt{\lambda\coth\lambda-\lambda^2},
\end{equation} for $\lambda$
in the domain $[0,\lambda_0]$ with $\lambda_0 \simeq 1.1997$ 
such that $\tanh\lambda_0 = 1/\lambda_0$. In terms of the 
``normalized--at--infinity'' outgoing solution\footnote{In 
the introduction of Paper II, the correspondence between 
$W_\ell(z)$ and $K_{\ell+1/2}(z)$ is off by a factor of 
$\pi/2$.} 
\begin{equation}
W_\ell(z) = \sqrt{\frac{2z}{\pi}}
\exp(z)K_{\ell+1/2}(z), 
\label{flatspaceW}
\end{equation}
we have
\begin{equation}
\hat{\omega}^{F}_\ell(\sigma;\rho_B) = 
\sigma\rho_B \frac{W_\ell'(\sigma\rho_B)}{W_\ell(\sigma\rho_B)}
\end{equation}
as another expression for the flatspace frequency--domain 
kernel \cite{AGH}. In passing, we remark that although the
exact FDNK (\ref{fdnrk}) is already a rational function, the
technique of kernel compression still proves useful for 
high--$\ell$ FDNK, since for a given tolerance $\varepsilon$ 
compression of (\ref{fdnrk}) yields a numerical kernel with
far fewer poles \cite{AGH}.

\subsection{Gravitational FDRK and outgoing solution}
Let us first consider the relationship between a gravitational 
FDRK $\hat{\omega}_\ell(\sigma;\rho_B)$ described earlier and 
solutions to the formal Laplace transform
\begin{equation}
\frac{\mathrm{d}^2\hat{\Psi}_\ell}{\mathrm{d}\rho_*^2}
- V(\rho)\hat{\Psi}_\ell = \sigma^2\hat{\Psi}_{\ell}
\label{FDwaveeq}
\end{equation}
of the generic equation (\ref{onedeewaveeq}). Here we mostly 
just collect relevant formulas. These formulas are derived in the 
first section of Paper I, and the derivation presented there goes 
through for Regge--Wheeler case ($\jmath = 0,1,2$ considered 
there) as well as the Zerilli case (not considered there). Whence 
the formulas we consider now are valid for all cases. 

We ``peel off'' the exponential behavior of the field by 
setting $\hat{\Psi}_\ell = \exp(-\sigma\rho_*)\hat{\Phi}_\ell$, 
whereupon finding 
\begin{equation}
                \frac{\mathrm{d}^2
                 \hat{\Phi}_\ell}{\mathrm{d}\rho^2}
                +\left(-2\sigma - \frac{1}{\rho}
                +\frac{1-2\sigma}{\rho-1} \right)
                 \frac{\mathrm{d}
                 \hat{\Phi}_\ell}{\mathrm{d}\rho}
                -\frac{\rho^2V(\rho)}{(\rho-1)^2}
                 \hat{\Phi}_\ell = 0
\label{maineq6}
\end{equation}
as the ODE satisfied by $\hat{\Phi}_\ell$. When numerically 
integrating (\ref{maineq6}) in various contexts, we have found it 
useful to work instead with $z = \sigma\rho$, re-expressing the 
equation as follows:
\begin{equation}
                \frac{\mathrm{d}^2
                 \hat{\Phi}_\ell}{\mathrm{d}z^2}
                +\left(-2 - \frac{1}{z}
                +\frac{1-2\sigma}{z-\sigma} \right)
                 \frac{\mathrm{d}
                 \hat{\Phi}_\ell}{\mathrm{d}z}
                -\frac{z^2 V(z/\sigma)}{
                 \sigma^2(z-\sigma)^2}
                 \hat{\Phi}_\ell = 0.
\label{maineq7}
\end{equation}
Let $W_\ell(z;\sigma)$ denote the outgoing solution to (\ref{maineq7}).
This solution obeys \cite{Olver}
\begin{equation}
W_\ell(z;\sigma) \sim \sum_{k=0}^\infty d_k(\sigma) z^{-k}
\label{Wasympexp}
\end{equation}
as $z\rightarrow \infty$, and we describe it as ``normalized at infinity.''
Appendix B considers the recursion relations defining the $d^{RW}_k(\sigma)$ and 
$d^{Z}_k(\sigma)$. As described in Paper I, the flatspace solution $W_\ell(z)$ 
in (\ref{flatspaceW}) is formally $W_\ell(z) = W^{RW}_\ell(z;0)$ in terms 
of the Regge--Wheeler case $W^{RW}_\ell(z;\sigma)$.

In our notation $W_\ell(\sigma\rho;\sigma)$ is the outgoing solution to 
(\ref{maineq6}), and Paper I expresses the FDRK in terms of it as
\begin{equation}
\hat{\omega}_\ell(\sigma;\rho_B)
= \sigma\rho_B \frac{W'_\ell(\sigma\rho_B;\sigma)}{W_\ell(\sigma\rho_B;\sigma)},
\label{FDRKkern}
\end{equation}
where the prime denotes differentiation in the first slot of 
$W_\ell(z;\sigma)$. Since we have peeled off the exponential factor 
$\exp(-\sigma\rho_*)$ above and now work with an outgoing solution
$W_\ell(\sigma\rho,\sigma)$ which is normalized at infinity, the 
FDRK (\ref{FDRKkern}) is more apt to have a well--defined inverse 
Laplace transform.

Paper I has described a collection of numerical methods which allow
us to (i) evaluate $W_\ell(\sigma\rho_B;\sigma)$ for 
$\sigma$ in the lefthalf plane and (ii) evaluate 
$\hat{\omega}_\ell(\mathrm{i}y;\rho_B)$ for real $y$. We remark that 
Leaver has analytically represented a solution, say the outgoing one 
$W_\ell^{RW}(\sigma\rho_B;\sigma)\exp(-\sigma\rho_*)$, to the 
frequency--domain Regge--Wheeler equation as an infinite series in 
Coulomb wavefunctions, 
where the expansion coefficients obey a three--term recursion relation. 
In fact, such series expansions hold more generally for the generalized 
spheroidal wave equation (essentially the confluent Heun equation) 
\cite{Leaver}. The methods described in Paper I are not based on the 
appropriate Leaver series, rather they rely on direct integration of
(\ref{maineq7}). As such they can and have 
now been carried over to the Zerilli case. We note that the Zerilli 
equation is not directly related to the confluent Heun equation; whence 
it is not immediately evident how to evaluate 
$W^Z_\ell(\sigma\rho_B;\sigma)$ via a Leaver series. 

Roughly, our method for evaluating $W_\ell(\sigma\rho_B;\sigma)$ is 
as follows. For a fixed frequency $\sigma$, initial data at a large
radius are obtained for (\ref{maineq7}) using the asymptotic expansion 
(\ref{Wasympexp}). Then (\ref{maineq7}) is integrated over a suitable 
path in the complex $z = \sigma\rho$ plane to a terminal point 
$z_B = \sigma\rho_B$. To convey the basic idea behind our numerical 
evaluation of the FDRK $\hat{\omega}_\ell(\mathrm{i}y;\rho_B)$ itself, 
we introduce
\begin{equation}
w_\ell(z;\sigma) = z\frac{W_\ell'(z;\sigma)}{W_\ell(z;\sigma)}.
\end{equation}
Then the FDRK is $\hat{\omega}_\ell(\sigma;\rho_B)
= w_\ell(\sigma\rho_B;\sigma)$, and we have from (\ref{maineq7}) 
that $w_\ell(z;\sigma)$ obeys the first--order nonlinear equation
\begin{equation}
\frac{\mathrm{d}w_\ell}{\mathrm{d}z}
+ \frac{w^2_\ell}{z}
+\left(-2 - \frac{2}{z}
+\frac{1-2\sigma}{z-\sigma} \right)w_\ell
-\frac{z^3 V(z/\sigma)}{
                 \sigma^2(z-\sigma)^2} = 0.
\label{1nonlinode}
\end{equation}
We calculate values 
$\hat{\omega}_\ell(\mathrm{i}y;\rho_B) =
w_\ell(\mathrm{i}y\rho_B;\mathrm{i}y)$ via direct numerical 
integration of (\ref{1nonlinode}). To achieve stability and high 
accuracy, the methods associated with both (i) and (ii) evaluations
require integration over nontrivial paths in the $z$--plane. 
Moreover, for technical reasons the integration employed for 
kernel evaluation (ii) is sometimes carried out in the complex 
$\rho$--plane rather than complex $z$--plane. Finally, we remark
that to produce the origin value $\hat{\omega}_\ell(0;\rho_B)$
of the kernel, we do not integrate (\ref{1nonlinode}). Rather,
we make use of an exact series expression, one given for the 
Regge--Wheeler cases in Paper I and for the Zerilli case in 
Appendix A.

\subsection{Gravitational sum--of--poles representation}
For the case of $\jmath = 0$ scalar perturbations Paper I has 
documented compelling numerical evidence indicating that 
the FDRK (\ref{FDRKkern}) admits an explicit ``sum--of--poles'' 
representation,
\begin{equation}
\hat{\omega}_\ell(\sigma;\rho_B) = \sum_{k = 1}^{N_\ell}
\frac{\alpha_{\ell,k}(\rho_B)}{\sigma-\sigma_{\ell,k}(\rho_B)} -
\frac{1}{\pi}\int^\infty_0
\frac{f_\ell(\chi;\rho_B)}{\sigma+\chi}\,{\rm d}\chi\, ,
\label{poleandcut}
\end{equation}
in terms of complex frequency $\sigma$. When we discussed 
compressed kernels before, we introduced approximate pole 
locations $\beta_{\ell,k}(\rho_B)$ and strengths 
$\gamma_{\ell,k}(\rho_B)$. We now consider $N_\ell$ (an integer) 
{\em physical pole locations} $\sigma_{\ell,k}(\rho_B)$ and
{\em physical pole strengths} $\alpha_{\ell,k}(\rho_B)$, 
all of which lie in the lefthalf plane. Also appearing in 
(\ref{poleandcut}) is a {\em cut profile} $f_\ell(\chi;\rho_B)$, 
and like the pole locations and strengths it depends on the value 
of $\rho_B$.\footnote{It may or may not be the case that an 
approximate location $\beta_{\ell,k}(\rho_B)$, for example, 
approximates a physical one $\sigma_{\ell,k}(\rho_B)$. It is 
the compressed kernel $\hat{\xi}_{\ell}(\mathrm{i}y;\rho_B)$ in 
whole which approximates the physical FDRK 
$\hat{\omega}_{\ell}(\mathrm{i}y;\rho_B)$ uniformly in 
$y\in\mathbb{R}$.} In principle the integer $N_\ell$ also 
depends $\rho_B$, but turns out to be constant over sizable 
regions of the relevant parameter space (more comments on this 
point below). The $k$th pole strength and cut profile are 
given respectively by
\begin{equation} \alpha_{\ell,k}(\rho_B) = -\rho_B
\sigma_{\ell,k}'(\rho_B)\, ,\quad f_\ell(\chi;\rho_B) =
\mathrm{Im}\hat{\omega}_\ell(\chi e^{{\rm i}\pi};\rho_B)\, ,
\end{equation}
with $\chi \geq 0$ and the prime here standing for 
$\partial/\partial\rho_B$ differentiation. 

We will argue that the representation (\ref{poleandcut}) is
also valid for both $\jmath = 2$ Regge--Wheeler and Zerilli 
gravitational cases, and this conjecture is the main result 
of our paper. Note that (\ref{poleandcut}) is not really a
``sum of poles.'' Indeed, recall that in the sense of complex 
analysis a pole is an {\em isolated} singularity. Strictly 
speaking then, the cut integral in (\ref{poleandcut}) does 
not correspond to a ``continuous distribution of poles'' (an 
oxymoron). Nevertheless, we shall continue to describe the 
representation (\ref{poleandcut}) as a ``sum of poles.''

We stress that (\ref{poleandcut}) is a representation for a 
{\em boundary} integral kernel. Its Laplace transform, the TDRK 
$\omega_\ell(\tau;\rho_B)$, lives on the history of the spatial boundary
$B$. Indeed, (\ref{therobc}) makes no reference to the details of the
initial data and is certainly not a spatial convolution over initial
data. Moreover, the pole locations $\sigma_{\ell,k}(\rho_B)$ are not
quasinormal modes. An infinite number of quasinormal modes belong to
each $\ell$ value, and these characteristic frequencies do not depend
on any particular choice of outer boundary radius $\rho_B$. Quasinormal 
modes are associated with a boundary value problem specifying that 
$\hat{\Psi}_\ell$ is downgoing at the horizon and outgoing at infinity. 
The locations $\sigma_{\ell,k}(\rho_B)$ are finite in number, and they do 
depend on $\rho_B$. They can be associated with a boundary value problem 
specifying that $\hat{\Psi}_\ell$ is outgoing at infinity and vanishes
at $\rho_B$.\footnote{As such, the locations $\sigma_{\ell,k}(\rho_B)$ 
are analogous to the ``flatspace quasinormal modes'' considered in
\cite{NollertSchmidt}, a misleading terminology for our paper.}
Likewise, the cut integral appearing in (\ref{poleandcut}) is not the
branch--cut contribution to the usual Green's function studied in
the quasinormal mode problem \cite{Andersson}.

\section{Numerical study}
In this section we both provide further qualitative description of 
the key representation (\ref{poleandcut}) and justify it numerically. 
To argue that the representation (\ref{poleandcut}) is valid for both 
$\jmath = 2$ Regge--Wheeler and Zerilli gravitational cases, we 
offer nearly the same evidence as that offered in Paper I for the 
case of scalar perturbations. However, here we also consider one extra 
numerical experiment based on the Argument Principle. Our main 
numerical justification is to compare values 
$\hat{\omega}_\ell(\mathrm{i}y;\rho_B)$ of the kernel
obtain via two independent approaches. These are the following:

\begin{itemize}
\item
Direct integration of (\ref{1nonlinode}) as alluded to above
[a process which makes no use whatsoever of the conjecture 
representation (\ref{poleandcut})]. 

\item
Approximation of the sum--of--poles representation itself
(by this we do {\em not} mean kernel compression).
\end{itemize}

In the second approach, we ``build'' the kernel as
\begin{eqnarray}
\hat{\omega}_\ell(\mathrm{i}y;\rho_B) & \simeq & 
\sum_{k = 1}^{N_\ell}
\frac{\alpha_{\ell,k}(\rho_B)
+\delta\alpha_{\ell,k}(\rho_B)}{\mathrm{i}y
-\sigma_{\ell,k}(\rho_B)-\delta\sigma_{\ell,k}(\rho_B)}
\label{imagpoleandcut} \\
& &  -
\frac{1}{\pi}\int^{\chi_\mathrm{max}}_{\chi_\mathrm{min}}
\frac{f_\ell(\chi;\rho_B)
+\delta f_\ell(\chi;\rho_B)}{\mathrm{i}y+\chi}\,{\rm d}\chi\, ,
\nonumber
\end{eqnarray}
where the $\delta$ terms represent numerical errors and the 
integral over the chosen window 
$[\chi_\mathrm{min},\chi_\mathrm{max}]$ must be handled via 
numerical quadrature. We have used Simpson's rule, respectively
with 2048, 2048, 1024, 2048, 1024, 2048, 1024, 2048, 512 
subintervals for $\ell =$ 2, 3, 4, 5, 6, 7, 8, 9, 10 (odd 
$\ell$ require more subintervals). We stress 
that this latter approach to evaluation, quite unlike the first 
approach, requires that we numerically compute all pole locations 
and strengths and well as the cut profile. To locate poles,
zeros in $\sigma$ of $W_\ell(\sigma\rho_B;\sigma)$, we have used 
the secant algorithm. The $\partial/\partial\rho_B$ derivatives
of the $\sigma_{\ell,k}(\rho_B)$ needed to compute the strengths
$\alpha_{\ell,k}(\rho_B)$ are obtained by first building a
high--order interpolating polynomial $T_{\ell,k}(1/\rho_B)$
for each $\sigma_{\ell,k}(\rho_B)$ based on Chebyshev nodes in 
$1/\rho_B$. Derivatives are then
found via differentiation of the Chebyshev polynomial. This
procedure is described in more detail in Paper I where it was
used for the $\jmath = 0$ Regge--Wheeler case.

We remark that this direct approximation
(\ref{imagpoleandcut}) to the sum--of--poles representation 
(\ref{poleandcut}) is certainly not a compressed kernel. 
Indeed, as mentioned, this brute--force approximation of 
(\ref{poleandcut}) typically requires thousands of poles (stemming 
from the numerical quadrature of the cut integral) to achieve the 
same error tolerance achieved by a compressed kernel 
comprised of ten or so poles.

Besides making the comparison outlined in the last paragraph, we also
wish to provide further qualitative description of the
sum--of--poles representation (\ref{poleandcut}). We describe both the
poles and the cut profile in more detail, and also compare the representation
to the strikingly similar representation (\ref{fdnrk}) of the FDNK 
$\hat{\omega}^{F}_\ell(\sigma;\rho_B)$ for a flatspace
order--$\ell$ multipole. We have found that both gravitational kernels, 
$\hat{\omega}^{RW}_\ell(\sigma;\rho_B)$ and 
$\hat{\omega}^{Z}_\ell(\sigma;\rho_B)$, indeed agree with 
$\hat{\omega}^{F}_\ell(\sigma;\rho_B)$ in the $\rho_B \rightarrow 
\infty$ limit, although here we will focus on the Zerilli case.
One certainly expects such agreement, since the Schwarzschild
solution is asymptotically flat. However, the nature of this asymptotic
agreement is rather interesting, and we point out some subtleties not
mentioned in Paper I.
\begin{figure}
\centering
\includegraphics[totalheight=0.35\textheight]{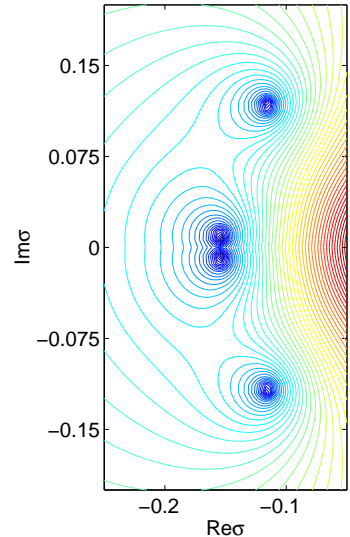}
\caption{Zeros $\{\sigma^Z_{3,k}(15): 1\leq k\leq 4\}$ in frequency of
$W_3^Z(\sigma 15;\sigma)$. The contour lines are of $\log_{10}
|W_3^Z(\sigma 15;\sigma)|$, with the logarithm distributing contour 
lines more evenly.}
\label{zerillizeros}
\end{figure}
\begin{figure}
\includegraphics[width=0.3\textwidth]{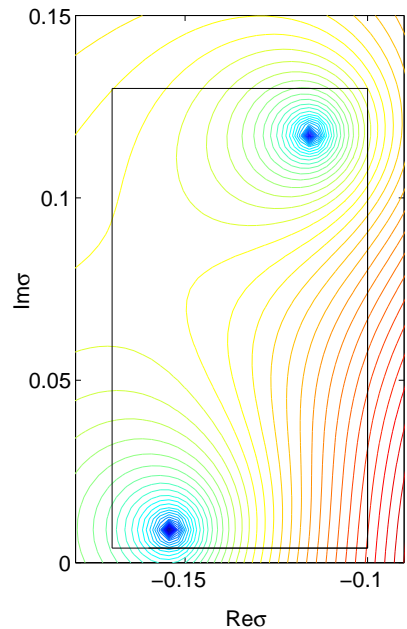}
\caption{Zeros in frequency of $W_3^Z(\sigma 15;\sigma)$. The plot
here is a blow--up of the one shown in {\sc FIG.}~\ref{zerillizeros}.}
\label{ell3rho15zerosZ}
\end{figure}

\subsection{Pole locations and strengths}
For the most part, this subsection gives what amounts to a
qualitative description. Using our method for evaluating the outgoing 
solution $W_\ell(\sigma \rho_B;\sigma)$, we may plot the modulus
$|W_\ell(\sigma \rho_B;\sigma)|$ in order to suggest rough 
values for zeros (that is, roots) of the function. Provided 
such a zero is simple, it will correspond to a pole appearing 
in the representation (\ref{poleandcut}). Over the parameter space
$\rho_B \geq 15$ and $\ell = 2,3,\cdots 10$, and for both Zerilli
and $\jmath = 2$ Regge--Wheeler cases, we have found that the
number $N_\ell$ of zeros $\sigma_{\ell,k}(\rho_B)$ is as
follows: $N_2 = 2$, $N_3 = 4 = N_4$, $N_5 = 6 = N_6$, $N_7 = 8
= N_8$, $N_9 = 10 = N_{10}$. For a fixed choice of $\ell$ and
$\rho_B$, these zeros form a crescent pattern in the lefthalf
$\sigma$--plane.

\begin{figure}
\includegraphics[width=0.3\textwidth]{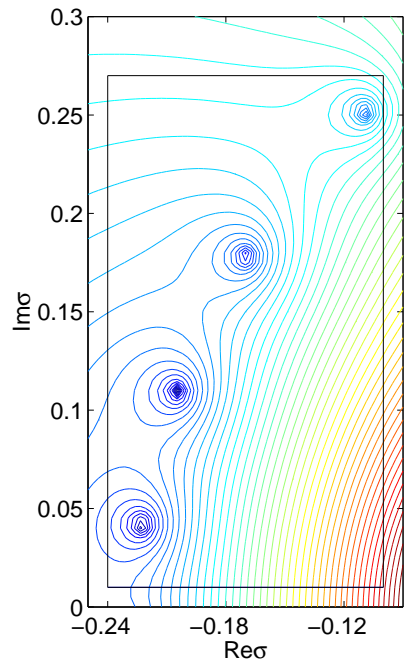}
\caption{Zeros in frequency of $W_8^Z(\sigma 25;\sigma)$.
The contour lines are of $\log_{10}|W_8^Z(\sigma 25;\sigma)|$, 
with the logarithm distributing contour lines more evenly.}
\label{ell8rho25zerosZ}
\end{figure}

For example, {\sc FIG.}~\ref{zerillizeros} depicts 
the modulus $|W_3^Z(\sigma 15;\sigma)|$ in the indicated region of 
the lefthalf $\sigma$ plane. Four zeros appear to be evident in 
the figure, and in order to further explore whether or not they
are indeed zeros, we appeal to the Argument Principle. We focus 
on the two upper locations shown closer up in 
{\sc FIG.}~\ref{ell3rho15zerosZ}. Let $h(\sigma) = 
W^Z_3(\sigma 15;\sigma)$ represent our numerically computed 
function.\footnote{Due to small errors in the asymptotic
expansion (\ref{Wasympexp}) used to generate initial data
for an evaluation based on integrating (\ref{maineq7}),
$h(\sigma)$ will actually represent the product of
$W^Z_3(\sigma 15;\sigma)$ and an analytic function of $\sigma$
which slowly varies over the region of interest. However,
Paper I showed that the choice of integration path in the
$z= \sigma\rho$ plane results in exponential suppression of
the second solution to (\ref{maineq7}), and absolute differences
in the zero {\em locations} of $h(\sigma)$ and those of 
$W^Z_3(\sigma 15;\sigma)$ are of size $10^{-13}$ in 
modulus.} We numerically compute
\begin{equation}
\frac{1}{2\pi\mathrm{i}}\oint_\mathrm{square}
\frac{h'(\sigma)}{h(\sigma)}\mathrm{d}\sigma
\simeq 1.999732-\mathrm{i}4.185506\times 10^{-5}
\end{equation}
over the square (running in the counterclockwise sense) 
shown in {\sc FIG.}~\ref{ell3rho15zerosZ}. On each side of the square
we have introduced 1024 subintervals, and used the 
trapezoid rule. Since we are unable to numerically evaluate 
$h'(\sigma)$ directly, we approximate this derivative using difference 
quotients (this requires two extra function evaluations beyond 
the corners). Using a sequence of discretizations, we have
confirmed that the integral convergences to $2$ at a second--order rate, 
suggesting that the square indeed encloses two simple
zeros. As a second example, consider the modulus 
$|W_8^Z(\sigma 25;\sigma)|$ plotted in {\sc FIG.}~\ref{ell8rho25zerosZ} 
over the indicated region (four other zeros, conjugate to those 
shown, are located in the third quadrant). For the analogous 
line--integral over the square shown (and using 2048 cells on
each side for the trapezoidal integration), we find a value of 
$3.999975-\mathrm{i}5.169761\times 10^{-5}$, suggesting
four simple zeros.
%
%
    
As with the scalar case, we find for the Zerilli and $\jmath = 2$ 
Regge--Wheeler cases that the zeros $\sigma_{\ell,k}(\rho_B)$ in 
frequency of $W_\ell(\sigma\rho_B;\sigma)$ behave asymptotically as 
\begin{equation}
\sigma_{\ell,k}(\rho_B) \sim b_{\ell,k}/\rho_B
\label{asympzeros}
\end{equation}
in the $\rho_B \rightarrow \infty$ limit. However, let us offer
several important observations in order to sharpen the precise nature 
of this asymptotic agreement.
First, in the opposite limit as $\rho_B \rightarrow 1^{+}$, 
we do not believe $N_\ell$ remains constant. Indeed, we expect the 
phenomenon of ``zero pair creation'' in this limit, as described in 
Paper I for the scalar $\jmath = 0$ Regge--Wheeler FDRK. 
Although we believe that the sum--of--poles representation remains 
valid for $\rho_B$ close (but not equal) to unity, description of
the FDRK and implementation of ROBC both become more difficult in
this limit. For these reasons, we have required $\rho_B \geq 15$. Second, 
for odd $\ell$ there is an ``extra'' zero. That is to say, for odd $\ell$
the number $N_\ell = \ell + 1$ of zeros $\sigma_{\ell,k}(\rho_B)$ 
is greater by one than the number $\ell$ of MacDonald zeros (provided 
$\rho_B$ is large enough, otherwise $N_\ell$ could be a larger 
integer still). We will argue that this curious feature is not at odds 
with the asymptotic result (\ref{asympzeros}) above. Let us focus 
on the Zerilli case, with the understanding that similar statements 
apply to the $\jmath = 2$ Regge--Wheeler case.
\begin{figure}
\vskip 5mm
\includegraphics[width=0.275\textwidth]{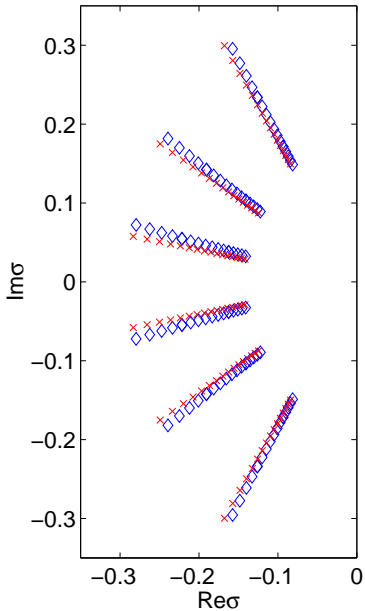}
\caption{Zeros $\{\sigma^Z_{6,k}(\rho_B): 1\leq k\leq 6\}$ in
frequency of $W_6^Z(\sigma \rho_B;\sigma)$. Diamonds represent the
zeros $\sigma_{6,k}(\rho_B)$, while the crosses represent
the $b_{6,k}/\rho_B$. All zeros for $\rho_B = 15,16,\cdots,30$ are
shown. The outermost crescent of locations corresponds
to $\rho_B = 15$ and the innermost to $\rho_B = 30$.}
\label{ZerrBessRootsEll6}
\end{figure}

For even $\ell = 2,4,6,8,10$ and $\rho_B \geq 15$, the number of
zeros $\sigma^Z_{\ell,k}(\rho_B)$ is $N_\ell = \ell$, the same as the
number of $b_{\ell,k}/\rho_B$. As an illustration, FIG.~\ref{ZerrBessRootsEll6}
depicts the six zeros $\sigma^Z_{6,k}(\rho_B)$ of $W_6^Z(\sigma \rho_B;\sigma)$ 
as diamonds for $\rho_B = 15,16,\cdots,30$. We also plot the corresponding 
MacDonald--Bessel zeros $b_{6,k}/\rho_B$ as crosses. The collection
$\sigma^Z_{6,k}(15)$ of zeros is the outermost crescent of diamonds, while
the collection $\sigma^Z_{6,k}(30)$ is the innermost (and similarly for the
crescents of Bessel crosses). The plot is clearly not at odds with the 
asymptotic formula (\ref{asympzeros}) above. 
\begin{figure}
\includegraphics[width=0.35\textwidth]{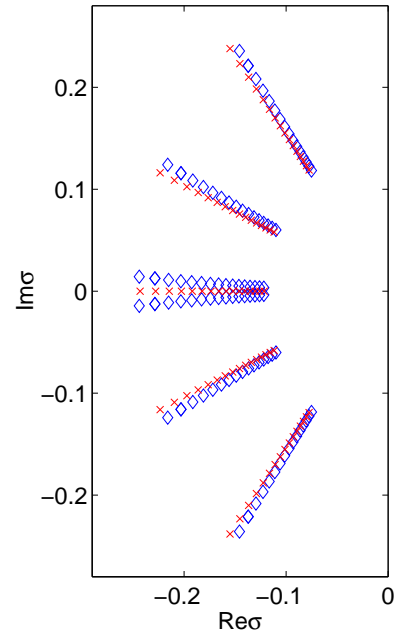}
\caption{Zeros $\{\sigma^Z_{5,k}(\rho_B): 1\leq k\leq 6\}$ in
frequency of $W_5^Z(\sigma \rho_B;\sigma)$. Diamonds represent the
zeros $\sigma^Z_{5,k}(\rho_B)$, while the crosses represent
$b_{5,k}/\rho_B$. All zeros for $\rho_B = 15,16,\cdots,30$ are
shown. The outermost crescent of locations corresponds
to $\rho_B = 15$ and the innermost to $\rho_B = 30$. For each
$\rho_B$ two diamonds correspond to the single cross lying on the
negative real axis.}
\label{ZerrBessRootsEll5}
\end{figure}

As mentioned, for odd $\ell = 1,3,5,7,9$ and $\rho_B \geq 15$, 
the number of zeros $\sigma^Z_{\ell,k}(\rho_B)$ is $N_\ell = \ell + 1$, 
that is one more than the corresponding number of MacDonald--Bessel 
zeros. For odd $\ell$ there is a single 
MacDonald--Bessel zero which lies on negative real axis. As with
the $\jmath = 0$ Regge--Wheeler case, we find for odd $\ell$ that
two zeros of $W^Z_\ell(\sigma\rho_B;\sigma)$ correspond to this
single real MacDonald--Bessel zero. Moreover, as $\rho_B$ gets large
each of these two zeros is asymptotic to the single MacDonald--Bessel 
zero. This phenomenon is evident in {\sc FIG.}~\ref{ZerrBessRootsEll5}, 
and corresponding plots for other odd $\ell$ are similar. The existence 
of  an ``extra'' zero is at first sight troubling in light of the 
expected asymptotic agreement between 
$\hat{\omega}^Z_\ell(\mathrm{i}y;\rho_B)$ and 
$\hat{\omega}^F_\ell(\mathrm{i}y;\rho_B)$. However, we argue below that
the pole and cut contributions to the gravitational FDRK in tandem
do yield the correct asymptotic agreement.

So far we have only consider locations $\sigma^Z_{\ell,k}(\rho_B)$
associated with Zerilli kernels. To the eye, both {\sc 
FIGS}.~\ref{ZerrBessRootsEll6} and \ref{ZerrBessRootsEll5} would be
the same had we instead plotted the corresponding locations
$\sigma^{RW}_{\ell,k}(\rho_B)$ associated with $\jmath = 2$ 
Regge--Wheeler kernels. For the examples we have considered,
$|\sigma^Z_{\ell,k}(\rho_B)-\sigma^{RW}_{\ell,k}(\rho_B)|$ is
typically on the order of $10^{-5}$ to $10^{-3}$. However, from
Paper I results we know that this difference is seven to ten
orders of magnitude larger than our knowledge of these locations.
As an example, we list the two conjugate locations 
$\sigma_{2,k}(15)$ for $\jmath = 0$ Regge--Wheeler, 
$\jmath = 2$ Regge--Wheeler, and Zerilli cases. Respectively, 
these are
\\[1mm]
$-0.096885391711329\pm\mathrm{i}0.061245961499841$,
\\$-0.094779501145744\pm\mathrm{i}0.059927941363806$,
\\$-0.093406539086840\pm\mathrm{i}0.058977077340679$.
\\[1mm]
These locations are all quite similar, and they may be
compared to the roots $b_{2,k}/15$: $-(3/2\pm\mathrm{i}
\sqrt{3}/2)/15 = -0.1\pm\mathrm{i}0.05773502691896$. In
this short list the last location, say, clearly corresponds in 
some sense to the ninth pole location in the compressed kernel in 
Table \ref{tab:tableZ}, but note that the absolute difference 
between the value here and the table value is more than the 
stated $\varepsilon = 10^{-10}$ tolerance for the compressed 
kernel. See the footnote just after Eq.~(\ref{poleandcut}).

\subsection{Cut profile}
\begin{figure}
\includegraphics[width=0.50\textwidth]{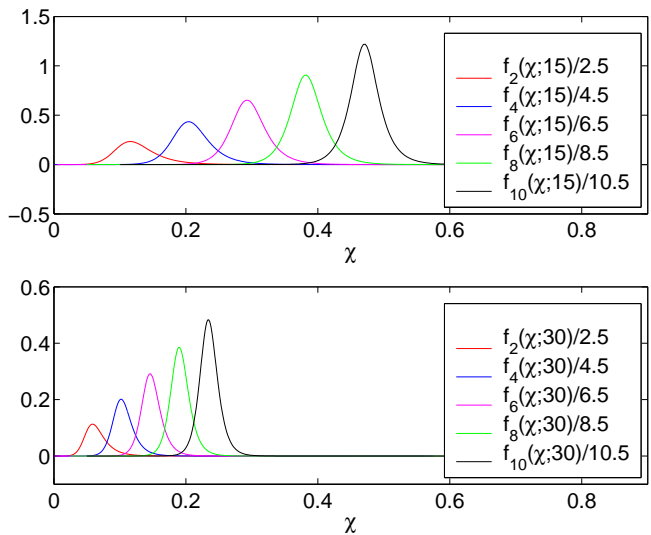}
\caption{Scaled even cut profiles for Zerilli kernels.
In each plot the leftmost profile is $f^Z_{2}(\chi;\rho_B)\big/2.5$
and the rightmost $f^Z_{10}(\chi;\rho_B)\big/10.5$.}
\label{ZerilliEvenCuts}
\end{figure}

\begin{figure}
\includegraphics[width=0.50\textwidth]{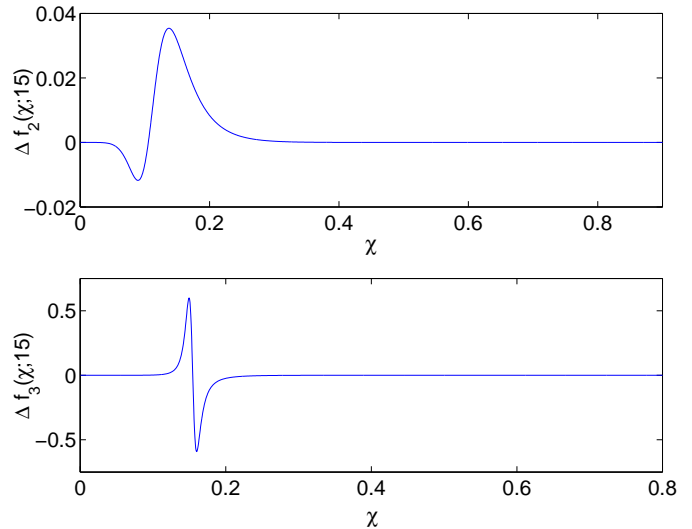}
\caption{{\sc Difference between Regge--Wheeler
and Zerilli cut profiles.} Here we plot example
differences with the notation $\Delta f_\ell(\chi;15) =
f^{RW}_\ell(\chi;15) - f^Z_\ell(\chi;15)$.}
\label{CompareCuts}
\end{figure}

Using the type (i) numerical method from Paper I described above
[which also returns derivative information for
$W_\ell'(\sigma\rho_B;\sigma)$], we have generated Zerilli cut 
profiles $f^Z_\ell(\chi;\rho_B)$ for various 
$\ell$ and $\rho_B$. {\sc FIG.}~\ref{ZerilliEvenCuts} depicts profiles
for even $\ell$. To the eye the cut profiles shown match 
$\jmath = 2$ Regge--Wheeler profiles $f^{RW}_\ell(\chi;\rho_B)$ 
corresponding to the same parameter choices. However, the Zerilli 
and Regge--Wheeler even profiles are indeed different, as is shown 
by an example plot in {\sc FIG.}~\ref{CompareCuts}. Near the peak of
this plotted difference, the accuracy to which we know the Zerilli 
and Regge--Wheeler profiles is some ten orders of magnitude greater 
than the difference. Notice that these even cut profiles {\em weaken} 
as $\rho_B$ gets larger, and we believe that the cut contribution
to an even--$\ell$ gravitational FDRK ``dies out'' as $\rho_B$ gets 
large. Based on our earlier claim that for an even--$\ell$ 
gravitational FDRK we have the same number $N_\ell = \ell$ of zeros 
$\sigma_{\ell,k}(\rho_B)$ as Bessel zeros $b_{\ell,k}/\rho_B$, and 
that these zeros lock on to the latter as $\rho_B$ gets large, we 
conjecture that the even--$\ell$ 
kernels $\hat{\omega}^Z_\ell(\mathrm{i}y;\rho_B)$
and $\hat{\omega}^{RW}_\ell(\mathrm{i}y;\rho_B)$ both 
approach $\hat{\omega}^F_\ell(\mathrm{i}y;\rho_B)$ uniformly in
$y$ as $\rho_B\rightarrow\infty$.
\begin{figure}
\includegraphics[width=0.50\textwidth]{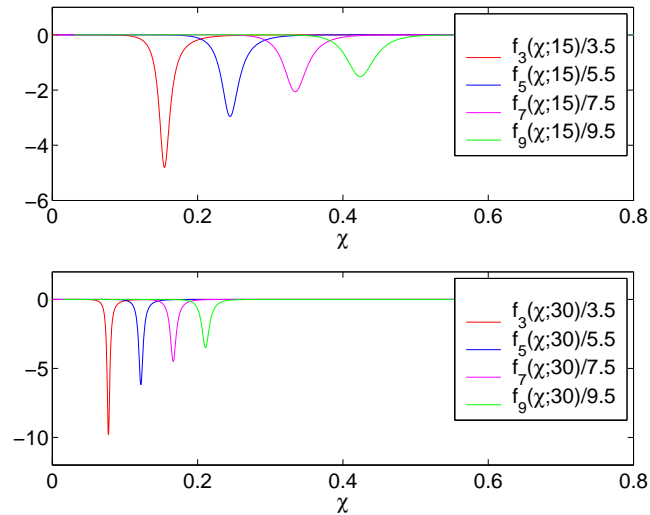}
\caption{Scaled odd cut profiles for Zerilli kernels.
In each plot the leftmost profile is $f^Z_{3}(\chi;\rho_B)\big/3.5$
and the rightmost $f^Z_{9}(\chi;\rho_B)\big/9.5$.}
\label{ZerilliOddCuts}
\end{figure}

We plot odd--$\ell$ Zerilli cut profiles in {\sc FIG.}~\ref{ZerilliOddCuts}. 
Like before with the even--$\ell$ profiles, to the eye these could be either 
Zerilli or $\jmath = 2$ Regge--Wheeler profiles. However, as is evident 
in the bottom plot shown in {\sc FIG.}~\ref{CompareCuts}, these cases are
different. Notice that for odd $\ell$ the cut profiles {\em strengthen} as 
$\rho_B$ increases. We believe that such a strengthening profile in tandem 
with the two zeros closest to the real axis as a whole combine to asymptotically 
agree with the single MacDonald--Bessel zero $b_{\ell,0}/\rho_B$ 
located on the negative real axis. [For notational simplicity we have now
labeled this zero by $k = 0$. Before it would have corresponded to
$k = \frac{1}{2}(\ell+1)$ for odd $\ell$, if $k = 1,\cdots,\ell$ for the
zeros $b_{\ell ,k}$ of $K_{\ell+1/2}(z)$.] In 
other words, for odd--$\ell$ the cut profile in effect cancels one
of the zeros $\sigma_{\ell,k}(\rho_B)$, so that we again we have agreement 
between the gravitational FDRK $\hat{\omega}^Z_\ell(\mathrm{i}y;\rho_B)$ and 
flatspace FDNK $\hat{\omega}^F_\ell(\mathrm{i}y;\rho_B)$ uniformly in $y$ 
as $\rho_B\rightarrow\infty$. While we cannot precisely describe how 
this cancellation of the extra zero occurs, we believe that it stems from
the following two conjectures, claimed to hold as $\rho_B \rightarrow
\infty$. First, the cut profile becomes sharply concentrated for $\chi$ 
near $-b_{\ell,0}/\rho_B$. Second,
\begin{equation}
-\frac{1}{\pi}\int_0^\infty f_\ell(\chi;\rho_B)\mathrm{d}\chi
\sim - b_{\ell,0}/\rho_B.
\end{equation}
Preliminary numerical investigations indicate that both claims are
in fact valid, but the issue deserves further study (preferably 
theoretical). All statements made in this paragraph also pertain to
$\hat{\omega}^{RW}_\ell(\mathrm{i}y;\rho_B)$. 

\subsection{Numerical validation of the 
            sum--of--pole representation}
\begin{figure}
\includegraphics[width=0.50\textwidth]{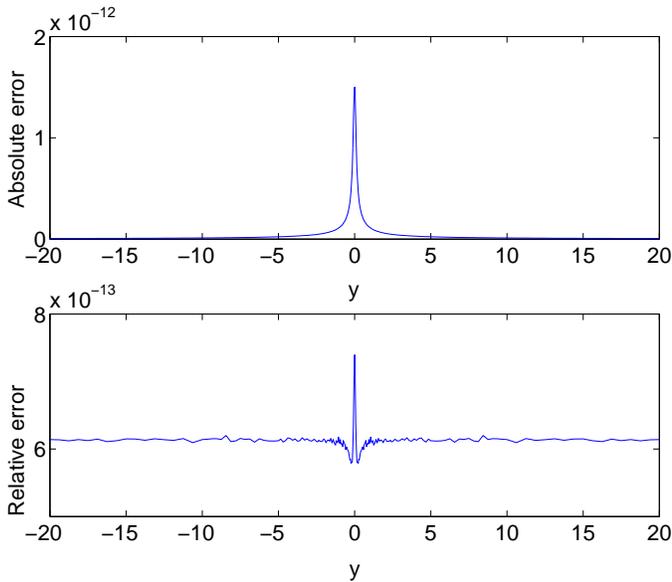}
\caption{{\sc Relative and absolute numerical errors.}
Here we plot numerical errors corresponding to
$|\Delta \hat{\omega}^Z_2(\mathrm{i}y;15)|$ and
$|\Delta \hat{\omega}^Z_2(\mathrm{i}y;15)|\big/
|\hat{\omega}^Z_2(\mathrm{i}y;15)|$
for the Zerilli case with $\ell = 2$ and $\rho_B = 15$.}
\label{ell2errors}
\end{figure}

Up to this point we have mainly given a qualitative description
of what we believe are the main features of the sum--of--poles
representation (\ref{poleandcut}). In order to quantitatively test 
our representation (\ref{poleandcut}), we compare the two numerical 
approaches for evaluating either 
$\hat{\omega}^Z_\ell(\mathrm{i}y;\rho_B)$ or 
$\hat{\omega}^{RW}_\ell(\mathrm{i}y;\rho_B)$ outlined at the 
beginning of this section. {\sc FIG.}~\ref{ell2errors} 
depicts such a comparison for the Zerilli case with $\ell = 2$ and 
$\rho_B = 15$. To make the plots in the figure, we have used a $y$--grid 
with 5 adaptive levels, each with 32 grid points. The adaptive 
grid provides more resolution near the origin where we 
expect the largest errors. We have then generated two 
separate numerical arrays of values 
$\hat{\omega}^Z_2(\mathrm{i}y;15)$ on the $y$--grid, the 
first obtained by integrating the ODE (\ref{1nonlinode}) 
and the second via the direct construction 
(\ref{imagpoleandcut}). Both arrays of numerical values 
are obtained in double precision arithmetic. The plots 
depict the absolute error 
$|\Delta \hat{\omega}^Z_2(\mathrm{i}y;15)|$ and 
relative error
$|\Delta \hat{\omega}^Z_2(\mathrm{i}y;15)|\big/
|\hat{\omega}^Z_2(\mathrm{i}y;15)|$ over the 
$y$--grid, and they indicate striking agreement between
the two methods. We note that the maximum value of the
absolute error in the top plot is $1.49\times 10^{-12}$,
and the maximum value of the relative error in the
bottom plot is $7.35 \times 10^{-13}$. We can now perform
the same experiment over a range of $\rho_B$ values, say
$\rho_B = 15,16,\cdots,30$. For each choice of $\rho_B$
we compute maximum values of absolute and relative error
over the $y$--grid. It turns out that the same values above
corresponding to $\rho_B = 15$ are the largest errors
encountered.

We now perform the same experiment for each $\ell \leq 10$ 
and for both $\jmath = 2$ Regge--Wheeler and Zerilli cases. That is
to say, for each $\ell$ we compute the maximum absolute error 
$|\Delta \hat{\omega}_\ell(\mathrm{i}y;\rho_B)|$ 
and maximum relative error 
$|\Delta \hat{\omega}_\ell(\mathrm{i}y;\rho_B)|\big/
|\hat{\omega}_\ell(\mathrm{i}y;\rho_B)|$ uniformly over the
described $y$--grid and all $\rho_B = 15, 16, \cdots, 30$. 
For the Zerilli case we list these errors in the Table 
\ref{zererrs}. Errors for the $\jmath = 2$ Regge--Wheeler case
are comparable. 
\begin{table}
\small
\begin{tabular}{lcc}
\hline
\\
$\ell$ & Max relative error & Max absolute error
 \\
\hline
2 & $7.35 \times 10^{-13}$ & $1.49\times 10^{-12}$ \\
\hline
3 & $3.63\times 10^{-12}$ & $9.50 \times 10^{-12}$ \\
\hline
4 & $8.63 \times 10^{-13}$ & $2.47\times 10^{-12}$ \\
\hline
5 & $1.27\times 10^{-12}$ & $4.68 \times 10^{-12}$ \\
\hline
6 & $6.17 \times 10^{-13}$ & $2.15\times 10^{-12}$ \\
\hline
7 & $6.93\times 10^{-13}$ & $4.69 \times 10^{-12}$ \\
\hline
8 & $3.03 \times 10^{-13}$ & $1.51\times 10^{-12}$ \\
\hline
9 & $5.10\times 10^{-13}$ & $3.51\times 10^{-12}$ \\
\hline
10 & $9.71 \times 10^{-14}$ & $8.92\times 10^{-13}$ \\
\hline
\end{tabular}
\caption{\sc Zerilli errors}
\label{zererrs}
\end{table}

\section{Conclusion}
For both flatspace 3+1 and flatspace 2+1 wave propagation and nonreflecting 
boundary conditions, AGH proved several theorems related to both the exact 
sum--of--poles representation for a FDNK (also a FDRK, but we have used FDNK
for the special flatspace case) and its numerical approximation \cite{AGH}. 
In particular, they proved that a FDNK admits a rational approximation as a 
compressed kernel, and for a given large Bessel order $\nu$ (for the 3+1 case
$\nu = \ell + 1/2$) and choice of $\varepsilon$ tolerance they estimated the
required number $d$ of poles appearing in the compressed kernel. In addition 
to some techniques stemming from the Fast Multipole Method, their proofs rely 
heavily on the well--understood theory of Bessel functions. In particular, 
AGH made extensive use of integral representations, order recursion relations, 
and detailed understanding of the scaling behavior for both the poles and (for 
the 2+1 case) the cut profile.\footnote{The 2+1 case, associated with 
integer--order Bessel functions, is analytically richer than the 3+1 case, as 
for the 2+1 case there is a continuous sector associated with the 
sum--of--poles representation for the FDNK. This sector is at least 
qualitatively similar to the one in (\ref{poleandcut}).} 

While the numerical evidenced amassed here is extremely convincing, it 
does not constitute a mathematical proof that a gravitational FDRK admits 
the representation (\ref{poleandcut}), and we hope that our numerical 
investigation spurs the interest of analysts capable of theoretically 
investigating our conjectured representation. Were (\ref{poleandcut})
established theoretically, we believe approximation theorems ---similar 
to those proved by AGH but pertaining to our gravitational ROBC--- would
follow. One might prove that a gravitational FDRK also admits a rational 
approximation as a compressed kernel, and determine the asymptotic 
growth of the number $d$ of approximating poles as $\ell \rightarrow 
\infty$, $\varepsilon \rightarrow 0^{+}$. Since numerically constructed 
compressed kernels have performed spectacularly in implementations of ROBC 
(see Paper II), we have good reason to believe that approximation theorems 
must hold. Such theorems are bound to involve the details of the underlying 
special functions. Unfortunately, relative to Bessel functions, significantly 
less is known about the special functions considered here, confluent Heun 
functions for the Regge--Wheeler cases and seemingly more exotic functions 
for the Zerilli case. Indeed, we are unaware of useful integral 
representations, and while appropriate Leaver series are certainly of formal 
interest, they would not seem a good platform for carrying out the requisite 
asymptotic analysis. While our own knowledge of modern analysis would seem 
not up to such theoretical investigation, we believe our results offer 
fertile new ground for more capable analysts.

\section{Acknowledgments}
I thank C.~R.~Evans, C.~O.~Lousto, and R.~H.~Price for clarifying remarks about 
blackhole perturbation theory, and gratefully acknowledge support from NSF grant 
PHY0514282.

\appendix

\section{Origin value of the FDRK} 
For the $\jmath = 0,1,2$ Regge--Wheeler cases Paper I has expressed the 
origin value $\hat{\omega}^{RW}_\ell(0;\rho_B)$ of the FDRK in terms of an 
infinite series in $\rho^{-1}_B$, where the expansion coefficients obey a 
two--term recursion relation (see Section 3.2.3 of that reference). All 
expansion coefficients are positive, and the value 
$\hat{\omega}^{RW}_\ell(0;\rho_B)$ can be accurately approximated 
in terms of two partial sums. This series was used numerically 
to evaluate the kernel at the origin $\sigma = 0$, a frequency value for 
which direct integration of (\ref{1nonlinode}) proved problematic.

To express $\hat{\omega}^Z_\ell(0;\rho_B)$ for the Zerilli case of polar 
perturbations, we have used a similar series expression, although now the
series coefficients obey a four--term rather than two--term recursion 
relation. We have obtained this series via the following recipe. 
First, we set $\sigma = 0$ in (\ref{maineq6}) and
use (\ref{Zerpotential}), thereby reaching
\begin{eqnarray}
                \lefteqn{\frac{\mathrm{d}^2
                 \hat{\Phi}_\ell}{\mathrm{d}\rho^2}
                +\left(- \frac{1}{\rho}
                +\frac{1}{\rho-1} \right)
                 \frac{\mathrm{d}
                 \hat{\Phi}_\ell}{\mathrm{d}\rho}} 
& & \nonumber \\
& &
                -\left[
                \frac{8n^2(n+1)\rho^3+ 12 n^2 + 18 n \rho 
                 + 9}{\rho^2(\rho-1)(2n\rho+3)^2}
                 \right]
                 \hat{\Phi}_\ell = 0.
\label{sigzero}
\end{eqnarray}
Next, plugging the expansion
\begin{equation}
\hat{\Phi}_\ell = \sum_{k=0}^\infty a_k\rho^{-(\ell+k)}
\end{equation}
into (\ref{sigzero}), assuming $a_0 = 1$, and balancing terms, 
we find the recursion relation
\begin{equation}
\beta_k a_{k+3} + \gamma_k a_{k+2} +\delta_k a_{k+1} + \epsilon_k
a_k = 0, 
\end{equation}
where
\begin{eqnarray}
\beta_k & = & 4n^2[(\ell+k+3)(\ell+k+4)-2(n+1)], \nonumber \\
\gamma_k & = & 4n(\ell+k+2)[(3-n)(\ell+k+3)-n]-12n^2,\nonumber \\
\delta_k & = & (\ell+k+1)[(9-12n)(\ell+k+2)-12n]-18n,\nonumber \\
\epsilon_k & = & -9(\ell+k+1)^2.
\end{eqnarray}
Again $n = \frac{1}{2}
(\ell - 1)(\ell + 2)$, and in 
the start--up, $\delta_{-2} = 0 = \epsilon_{-2}$ and $\epsilon_{-1} = 0$.
The origin value of the FDRK is
\begin{equation}
\hat{\omega}^Z_\ell(0;\rho_B) = 
\left.-\sum_{k=0}^\infty(\ell+k) a_k
\rho_B^{-(\ell+k)}\right/
\sum_{k=0}^\infty
a_k\rho_B^{-(\ell+k)}.
\label{origin}
\end{equation}
Despite the four--term recursion relation used to generate the series
and its $\rho$--derivative, in using (\ref{origin}) we have encountered
no numerical instabilities.

\section{Asymptotic expansion for outgoing solution}
To generate initial data for our numerical methods based on path
integration in the complex $z$--plane (or sometimes the complex
$\rho$--plane), we use the asymptotic expansion (\ref{Wasympexp})
about the irregular singular point at infinity.  Paper I describes
this expansion for all the Regge--Wheeler cases, showing that the
$d^{RW}_k(\sigma)$ obey a three--term recursion relation. For the 
Zerilli case, we write the expansion coefficients as $d^Z_k(\sigma) 
= g_k(\sigma) \sigma^k$, then finding that the $g_k(\sigma)$ obey 
the five--term recursion relation
\begin{equation}
A_k g_{k+4} + B_k g_{k+3} + C_k g_{k+2} + D_k
g_{k+1} + E_k g_k = 0,
\label{fiveterm}
\end{equation}
where
\begin{eqnarray}
A_k & = & 8\sigma n^2(k+4),\nonumber \\
B_k & = & 4n(k+3)[6\sigma + n(k+4)]-8n^2(n+1)\nonumber \\
C_k & = & (k+2)[18\sigma - 4n^2 + (12 n - 4n^2)(k+3)]-12n^2\nonumber \\
D_k & = & (k+1)[(9-12n)(k+2)-12n]-18n\nonumber \\
E_k & = & -9(k+1)^2.
\label{Wcoeffs}
\end{eqnarray}
Here again $n = \frac{1}{2}
(\ell - 1)(\ell + 2)$, and in the
start--up $C_{-3} = D_{-3} = E_{-3} = 0$,
$D_{-2} = E_{-2} = 0$, and $E_{-1} = 0$. Typically, we
have used fewer than ten terms in the expansion (\ref{Wasympexp}) 
to generate initial data for numerical integration,
and have encountered no problems in using this expansion
(despite potentially tricky issues associated
with high--order recursions). We have used dimensionless 
coordinates and Laplace rather than Fourier transform.
Adjusting for these choices, (\ref{fiveterm}) and 
(\ref{Wcoeffs}) agree with an expansion given by 
Chandrasekhar and Detweiler in \cite{ChandraDet2}.


\begin{thebibliography}{99}

\bibitem{Wheeler} J.~A.~Wheeler, 
Phys.~Rev.~{\bf 97}, (1955) 511.

\bibitem{ReggeWheeler} T.~Regge and J.~A.~Wheeler,
Phys.~Rev.~{\bf 108}, (1957) 1063.

\bibitem{Zerilli} F.~J.~Zerilli,
Phys.~Rev.~D {\bf 2}, (1970) 2141.

\bibitem{Price72b}
R.~H.~Price, Phys.~Rev.~D {\bf 5}, (1972) 2439.

\bibitem{ChandraDet} S.~Chandrasekhar, {\em
Mathematical Theory of Black Holes}, (Oxford 
University Press, Oxford, 1992).

\bibitem{AndPrice} A.~Anderson and R.~Price, 
Phys.~Rev.~D {\bf 43}, (1991) 3147--3154.

\bibitem{PricePullin}
R.~H.~Price and J.~Pullin, Phys.~Rev.~Lett.~{\bf 72}
(1994) 3297.

\bibitem{NBPP} H.--P.~Nollert, J.~Baker, R.~Price, and J.~Pullin, 
in Proceedings of the 18th Texas Symposium on 
Relativistic Astrophysics, {\tt gr-qc/9710011}.

\bibitem{GNPP1}
R.~Gleiser, O.~Nicasio, R.~Price, and J.~Pullin, 
Phys.~Rev. Lett.~{\bf 77} (1996) 4483.

\bibitem{GNPP2}
R.~Gleiser, O.~Nicasio, R.~Price, and J.~Pullin,
Phys.~Rev. D {\bf 59} (1999) 044024.

\bibitem{Lousto1} C.~O.~Lousto, Phys.~Rev.~Lett.~{\bf 84}
(2000) 5251.

\bibitem{Lousto2} C.~O.~Lousto, 
{\em A fourth order convergent numerical algorithm to integrate
nonrotating binary black hole perturbations in the extreme
mass ratio limit}, {\tt gr-qc/0503001}.

\bibitem{Baez} See {\tt http://math.ucr.edu/home/baez/area.html}
for a detailed literature review of this issue.

\bibitem{Olaf1} O.~Dreyer, Phys.~Rev.~Lett.~{\bf 90} (2003) 081301;
Contribution to the Proceedings of the 3rd International Symposium 
on Quantum Theory and Symmetries, Cincinnati, September 2003, 
{\tt gr-qc/0404055}.

\bibitem{DomLew} M.~Domagala and J.~Lewandowski, Class.~Quantum
Grav.~{\bf 21} (2004) 5233.

\bibitem{MotlandNeitzke} L.~Motl and A.~Neitzke,
Adv.~Theor.~Math.~Phys.~{\bf 7} (2003) 307.

\bibitem{MusiriSiopsis}
S.~Musiri and G.~Siopsis, Class.~Quant.~Grav.~{\bf 20} 
(2003) L285.

\bibitem{YorkSchmekel}
J.~W.~York, Jr. and B.~S.~Schmekel,
Phys.~Rev.~D {\bf 72}, (2005) 024022.
 
\bibitem{York} J.~W.~York, Jr.,
Phys.~Rev.~D {\bf 28} (1983) 2929.

\bibitem{AGH}
 B.~Alpert, L.~Greengard, and T.~Hagstrom,
SIAM J.~Numer.~Anal.~{\bf 37} (2000) 1138.

\bibitem{LauROBC1}
S.~R.~Lau, J.~Comput.~Phys.~{\bf 199} (2004) 376.

\bibitem{LauROBC2} S.~R.~Lau, 
Class.~and Quantum Grav~{\bf 21} (2004) 4147.

\bibitem{LauEvans}  C.~R.~Evans and S.~R.~Lau, 
{\em The tail of the outer boundary}, in preparation,
Summer and Fall 2005.

\bibitem{KokSchmidt}
K.~D.~Kokkotas and B.~G.~Schmidt,
``Quasi--Normal Modes of Stars and Black Holes,''
Living Reviews of Relativity {\bf 2}, 2, and
gr--qc/9909058.

\bibitem{CHE1} {\em Heun's Differential Equations}, edited by
A.~Ronveaux (Oxford University Press, Oxford, 1995).
                                                                                
\bibitem{CHE2} S.~Y.~Slavyanov and W.~Lay, {\em Special Functions:
a Unified Theory Based on Singularities} (Oxford University
Press, Oxford, 2000).

\bibitem{LeVeque}
R.~J.~LeVeque,
{\em Numerical Methods for Conservation Laws}, second edition
(Birkh\"{a}user Verlag, Basel, 1992).

\bibitem{Olver} Olver, F.~W.~J.~{\em Asymptotics and Special Functions}
(Academic Press, New York and London, 1974).

\bibitem{Leaver} E.~W.~Leaver, J.~Math.~Phys.~{\bf 27}
(1986) 1238.

\bibitem{NollertSchmidt}
H.--P.~Nollert and B.~G.~Schmidt,
Phys.~Rev.~D {\bf 45} (1992) 2617.

\bibitem{Andersson}  See, for example,
N.~Andersson, Phys.~Rev.~D {\bf 55} (1997) 468.

\bibitem{Watson} G.~N.~Watson, {\em A Treatise on the Theory
of Bessel Functions},
second edition (Cambridge University Press,
Cambridge, 1944).

\bibitem{Jiang} S.~Jiang, {\em Fast Evaluation of Nonreflecting
Boundary Conditions for the Schr\"{o}\-ding\-er Equation},
New York University Ph.D.~Dissertation (2001).

\bibitem{ChandraDet2} S.~Chandrasekhar and S.~Detweiler,
Proc.~R.~Soc.~Lond. A {\bf 344} (1975) 441.

\end{thebibliography}
\end{document}